\documentclass[aps,prl,twocolumn,groupedaddress]{revtex4-2}

\usepackage{amsmath, slashed, amssymb}
\def\K{{\mathcal{K}}}
\def\swedge{{\, \scriptstyle \wedge \,}}
\begin{document}


\title{Asymptotic gravitational charges}

\author{Hadi Godazgar}
\email[]{hadi.godazgar@aei.mpg.de}
\affiliation{Max-Planck-Institut f\"ur Gravitationsphysik (Albert-Einstein-Institut), M\"uhlenberg 1, D-14476 Potsdam, Germany.}
\author{Mahdi Godazgar}
\email[]{m.godazgar@qmul.ac.uk}
\affiliation{School of Mathematical Sciences, Queen Mary University of London, Mile End Road, London E1 4NS, UK.}
\author{Malcolm J. Perry}
\email{malcolm@damtp.cam.ac.uk}
\affiliation{School of Physics and Astronomy, Queen Mary University of London, Mile End Road, London E1 4NS, UK}
\altaffiliation{Also at Centre for Mathematical Sciences, Wilberforce Road, Cambridge, CB3 0WA, UK}
\altaffiliation{and Trinity College, Cambridge, CB2 1TQ, UK.}

\date{September 3, 2020}

\begin{abstract}
We present a method for finding, in principle, all asymptotic gravitational charges.  The basic idea is that one must consider all possible contributions to the action that do not affect the equations of motion for the theory of interest; such terms include topological terms.  As a result we observe that the first order formalism is best suited to an analysis of asymptotic charges.  In particular, this method can be used to provide a Hamiltonian derivation of recently found dual charges.
\end{abstract}


\maketitle

Symmetries are at the heart of our present understanding of fundamental physics. In gravitation, co-ordinate
invariance is a symmetry. If one includes fermionic matter, one needs to introduce, in addition to the metric,
the frame fields (or vierbeins) 
and then local Lorentz transformations are also 
symmetries. Some symmetries can be associated with charges as a consequence of Noether's theorem. 
A simple example of this was explored by Arnowitt, Deser and Misner
~\cite{Arnowitt:1961zz}.
A time translation diffeomorphism was shown to be associated to the total mass, the ADM mass, as measured at spatial 
infinity in an asymptotically flat spacetime. One might therefore expect that similar reasoning 
would produce charges associated with any of the generators of Poincar\'e transformations. What is
perhaps surprising, is that at null infinity, as was discovered by Bondi, van der Burg, Metzner and Sachs \cite{bondi, sachs}, there are an infinite
number of these asymptotic symmetries, BMS symmetries, that lead to an infinite number of physically meaningful charges, BMS charges. Since BMS charges are defined at null infinity, they are not exactly conserved like the ADM mass but satisfy a continuity equation. In other words they measure, and are sensitive to, the flux that is radiated away. For example, one important BMS charge is the Bondi mass, which is the quantity measured in gravitational wave observations.   

Recent
work has highlighted the importance of BMS charges in the computation of scattering amplitudes in processes
that involve massless particles~\cite{Strom:soft1}, the physics of gravitational waves and 
their detection \cite{Strominger:2014pwa} and in the black hole
information paradox~\cite{Hawking:2016msc}. The purpose of this letter is to systematically explore what these charges are for
gravitation.  Recently, it has been shown that, besides BMS charges, there are other asymptotic charges, dual charges, that encode the topology of spacetime \cite{dual0, dualex}. The origin of these charges has been hitherto not clear.  We argue that these asymptotic charges, in addition to the BMS charges, arise from different terms in the action that do not contribute to the equations of motion. For example, if we are interested in vacuum Einstein theory, by simply considering the Einstein-Hilbert action we miss out on dual charges. 
Therefore, we must consider all possible actions that give rise to the same equations of motion.  In applying this idea, in addition to finding the well-known BMS charges \cite{Dray:1984rfa, IW}, we give a Hamiltonian derivation of the recently found dual charges \cite{dual0, dualex}, and by corollary a Hamiltonian derivation of Newman-Penrose charges \cite{NP}, and show how other charges can be 
found from other topological contributions to the action---the physical significance of these will be explored in other work. 

The idea that topological terms can give rise to charges can be easily seen in electromagnetism. It is well-known that applying Noether's theorem to the Maxwell theory coupled to matter gives the electric charge. However, one can also add the $\theta$-term,
\begin{equation*}
  \int F \swedge F = -\frac{1}{4} \int \varepsilon\ \varepsilon^{\mu \nu \rho \sigma} F_{\mu \nu} F_{\rho \sigma},
\end{equation*}
to the theory and notice that its inclusion will lead to the magnetic charges. In this letter, we advocate an analogous approach in gravity.

We use the covariant phase space method and apply it to the 
BMS symmetries appropriate to asymptotically flat spacetimes, but we emphasise that this can be done generally for any 2-surface embedded in a three-dimensional space (see \cite{Jacobson:2015uqa, Frodden:2017qwh, Oliveri:2019gvm} for related considerations in the first-order formalism motivated mainly by the first law of black hole mechanics and \cite{Aros:1999id} for the study of topological actions in relation to charges).

\section{The Theory}

We will use the first order formalism of general relativity coupled to a Dirac field to illustrate the construction
of the gravitational charges. The first order formalism results in simpler expressions than the usual Einstein-
Hilbert formalism. Furthermore, because we are including the Dirac field, we find that torsion plays a significant
r\^ole~\cite{Kibble:1961ba}. There are three components to the total action.
The first is the Palatini term
\begin{equation}
I_P = \frac{1}{16\pi}\int_{\cal M} {\textstyle \frac{1}{2}} \, \varepsilon_{abcd}\, \mathcal{R}^{ab}(\omega) \swedge e^c \swedge e^d
\end{equation}
where $\mathcal{R}^{ab}(\omega)$ is the curvature 2-form made from the connection 1-form $\omega^{ab}$ by
$\mathcal{R}^{ab} = d\omega^{ab} + \omega^{ac}\swedge \omega_c{}^b.$
Lorentz indices $(a,b,\ldots)$ are lowered and raised using the
flat tangent space metric $\eta_{ab}$ and its inverse $\eta^{ab}$. Similarly, spacetime indices
$(\mu,\nu,\ldots)$ are lowered and raised
using the spacetime metric $g_{\mu\nu}$ and its inverse $g^{\mu\nu}$. 
$e^a$ are a pseudo-orthonormal basis of 1-forms such that the spacetime line element  $ds^2 = g_{\mu\nu}dx^\mu dx^\nu = \eta_{ab} e^a e^b$, thus
$e^a = e^a_\mu dx^\mu$ where $e^a_\mu$ are the components of the vierbeins and $d$ is the exterior 
derivative operator.
The metric connection $\omega^{ab}=\omega^{[ab]}$ and the torsion 2-form $T^a$ are defined by
$de^a + \omega^a{}_b\swedge e^b = T^a.$ In the first order formalism, one regards $e^a$ and $\omega^{ab}$
as independent variables. If $I_P$ were the only contribution to the action, then the equations
of motion would lead to the vacuum Einstein equation $R_{ab} = 0$ and vanishing of the torsion $T^a = 0.$
A second contribution to the action comes from anticommuting
Dirac fermions $\psi$. The Dirac action is
\begin{equation}
I_D = \int_{\cal M} \varepsilon\, ({\textstyle \frac{1}{2}}\bar\psi\gamma^a\nabla_a\psi -{\textstyle \frac{1}{2}}\nabla_a\bar\psi\gamma^a\psi -m\bar\psi\psi)
\end{equation}
where the volume form $\varepsilon = \frac{1}{24}\varepsilon_{abcd}e^a\swedge e^b\swedge e^c\swedge e^d.$
We define the Dirac conjugate to be $\bar\psi=i\psi^\dag\gamma^0$ and our
gamma matrix conventions are $\{\gamma^a,\gamma^b\}=2\eta^{ab}$ with the
signature being $(-+++)$. $\nabla_a\psi$ is the covariant
derivative given explicitly $\nabla_a\psi = \partial_a\psi + \frac{1}{4}\omega_a{}^{bc}\gamma_{bc}\psi$ with
$\omega^{ab}=\omega_c{}^{ab} e^c.$
The last contribution to the action is topological in nature and as a
consequence makes no contribution to the equations of
motion. It is in some sense a gravitational analog of the
Pontryagin index.
\begin{equation}
I_{NY}= \frac{i\lambda}{16\pi} \int_{\cal M} (\mathcal{R}_{ab}(\omega)\swedge e^a\swedge e^b - T^a\swedge T_a).
\end{equation}
The integrand is known as the Nieh-Yan tensor and is
exact, being given by $-d(e^a\swedge T_a)$. The factor of $i$ arises
because in a space of Euclidean signature, one would expect this
term to be real. However, in continuing to a Lorentzian
signatured spacetime, a factor of $i$ arises just as it does
for the Pontryagin term in Yang-Mills theory.
Equations of motion come from varying $e^a , \omega_{ab}$ and
$\psi$ in the total action $I_T = I_P + I_D + I_{NY}.$  Varying $\psi$
gives the Dirac equation $(\gamma^a \nabla_a-m)\psi=0.$ Varying $\omega^{ab}$ determines the torsion 
$T^a  =-2\pi \, \bar\psi\gamma^a{}_{bc}\psi \, e^b \swedge e^c.$
Thus only the totally antisymmetric part of the torsion is non-zero and is proportional
to the axial current of the fermion. Varying $e^{a}$ gives the
Einstein equation
\begin{equation}
R_{ab}-{\textstyle \frac{1}{2}} \eta_{ab} R = -4\pi(\bar\psi\gamma_a\nabla_b\psi -  \nabla_b\bar\psi\gamma_a\psi ).
\end{equation}

It should be noted that the Einstein equation is not
symmetric under the interchange of $a$ and $b$ when torsion is present. One could
write the the symmetric part as the conventional Einstein
equation coupled to the conventional symmetric energy-momentum tensor of the Dirac field. The antisymmetric
part is then a trivial consequence of the torsion equation
of motion and the Bianchi identity.

\section{Presymplectic potential and Noether charge}

The system $I_T$ admits two kinds of local invariance.
The first is diffeomorphism invariance which is a property
of all gravitational theories. The second is local Lorentz
invariance. The latter is necessitated because we have included 
fermions in our description of basic physics. The
first step is the application of Noether's theorem to find a
set of conserved currents or their dual $3$-forms. Once one
has found the currents, if the background field equations
are satisfied, the currents are conserved. Then one can
find an antisymmetric two-indexed tensor or its dual $2$-
form which can be integrated over a closed $2$-surface to
give the Noether charge on that surface. Typically, that
surface will be a sphere at infinity and the symmetry generator 
does not die off at infinity. This construction gives
rise to a charge for each of the asymptotic symmetries.
The value of the coefficient of the topological term $\lambda$ has
no effect on the dynamics of the theory. Consequently,
one can consider the symmetries coming from $I_{NY}$ to be
independent of those derived from $I_P + I_D.$ The charges
coming from $I_{NY}$ we will refer to as magnetic and those
from $I_P + I_D$ as electric. Any (smooth) vector field
$\xi^\mu$  can be used to generate an infinitesimal diffeomorphism and the actions
on $e^a, ~\omega_{ab}$ and $\psi$ are given by
\begin{equation*}
\delta_\xi e^a  = \mathcal{L}_\xi e^a, \quad 
\delta_\xi \omega_{ab}  = \mathcal{L}_\xi \omega_{ab}, \quad
\delta_\xi \psi  = \xi^\mu\partial_\mu\psi,
\end{equation*} 
where $\mathcal{L}_\xi$ is the Lie derivative with respect to the vector field $\xi.$
Infinitesimal Lorentz transformations are given by an antisymmetric two-indexed field
$\Lambda_{ab}$ and its action is
\begin{eqnarray*}
&\delta_\Lambda e^a =  \Lambda^a{}_b e^b,  \quad 
\delta_\Lambda\omega_{ab} = -d\Lambda_{ab} + [\Lambda,\omega]_{ab}, \\
& \delta_\Lambda \psi = \frac{1}{4}\Lambda_{ab}\gamma^{ab}\psi.
\end{eqnarray*}
We derive these charges and their properties using covariant phase space 
methods. The presymplectic potential $3$-form  $\theta$ is the boundary term found when the 
Lagrangian 4-form, $L$, is varied,
\begin{equation*}
 \delta L = E \delta \phi + d \theta,
\end{equation*}
where $E=0$ is the equation of motion and $\phi$ represents the fields.  

The electric contribution to $\theta$ is
\begin{align}
\theta_E = &\frac{1}{32\pi}\varepsilon_{abcd}\delta\omega^{ab} \swedge e^c\swedge e^d \notag \\
&-\frac{i}{12}(\bar\psi\gamma_{abc}\gamma_5\delta\psi-\delta\bar\psi\gamma_{abc}\gamma_5\psi)
e^a\swedge e^b\swedge e^c.
\end{align}
The magnetic contribution to $\theta$ is
\begin{equation}
\theta_M = \frac{i}{16\pi}(\delta \omega_{ab} \swedge e^a \swedge e^b -2\, \delta e^a \swedge T_a).
\end{equation}
The Noether currents are then given by
\begin{equation}
J_{\xi,\Lambda} = \theta(\delta_{\xi,\Lambda}\phi) - \iota_\xi  L,
\end{equation}
i.e.\
$\theta$ is evaluated with the variation relevant to the coordinate transformations generated by vector 
field $\xi$ and Lorentz transformations parametrised by $\Lambda$
in question.  
\subsection{Electric Noether charges}
When the equations of motion are satisfied, $J$ becomes the derivative of a $2$-form 
\begin{equation}
Q_E = \frac{1}{32\pi}\varepsilon_{abcd} \left(\iota_\xi\omega^{ab} - \Lambda^{ab} \right) e^c\swedge e^d.
\end{equation}

\subsection{Magnetic Noether charges}
The magnetic charges are in many ways similar:
\begin{equation}
Q_M=\frac{i}{16\pi}\left(\iota_\xi \omega^{ab} - \Lambda^{ab} \right) e_a\swedge e_b -\frac{1}{8\pi}(\iota_\xi e^a)T_a.
\end{equation}

\section{Variation of the Charges}

Each of the Noether charges is defined for a specific gauge transformation
and background. A problem is that the charge defined this way has no absolute physical
meaning as one could always add an arbitrary constant to the charge. What does have physical meaning
is to consider the change in charge conjugate to some specific transformation as one varies the background.
Let $\phi$ be the collection of fields $e^a,\omega^{ab}$ and $\psi$. Then we need to find the difference 
in a specific charge between $\phi$ and its variation $\phi + \delta\phi$.
The variation of a charge is constructed from the symplectic form $\Omega$ which is 
defined to be
\begin{equation}
\Omega = \int_{\Sigma} \left\{ \delta \theta(\phi,\delta^\prime\phi)
-\delta^\prime\theta(\phi,\delta\phi) \right\}
\end{equation}
where $\Sigma$ is a spacelike surface with boundary $\partial\Sigma$.  
If $\delta^\prime$ is chosen to be a gauge transformation, $\delta\phi$ obeys the linearised 
equations of motion and $\phi$ obeys the equations of motion, then $\Omega$ reduces to an
integral over  $\partial\Sigma$ and is the variation of the physical charge $\slashed\delta \mathcal{Q}$.
For any combination of diffeomorphisms and Lorentz transformations, 
\begin{eqnarray}
\slashed\delta \mathcal{Q} &=&   \int_{\partial\Sigma} \left( \delta Q -\iota_\xi \theta(\phi,\delta\phi) \right). \label{ham}
\end{eqnarray}
We have written the variation as $\slashed\delta \mathcal{Q}$ to indicate that the variation may not be exact.
The variation is supposed to measure what happens as one carries out the variation in a fixed region of spacetime.
The result should then be the change in the  physical charge and reflects the nature of the region in question.
However,
when carrying out the variation, some of the charge may have escaped through $\partial\Sigma$ and it is
this that leads to 
$\slashed\delta \mathcal{Q}$ not being exact. To find the exact piece, remove from $\slashed\delta \mathcal{Q}$
the piece that is not exact. Unfortunately, this prescription has some ambiguity as has been discussed in 
detail and partially resolved by Wald and Zoupas \cite{WZ}; see also \cite{Compere}. 
It is usually possible to understand the physics of
this process by finding a flux formula for the charge 
through $\partial\Sigma$.

\section{Asymptotic Evaluation}
One area that has been extensively explored is the evaluation of these charges at future null infinity.
Null infinity is a large sphere parametrised by the retarded time. The metric on null infinity is degenerate.
The approach to null infinity is carried out by taking the limit as a radial coordinate $r$ tends to infinity.  
This is often done in the Bondi gauge where the spacetime line element is of the form
\begin{equation*}
-Fe^{2\beta}du^2 -2e^{2\beta}dudr + r^2 h_{{\scriptscriptstyle IJ}}(dx^{\scriptscriptstyle I}-C^{\scriptscriptstyle I} du)(dx^{\scriptscriptstyle J} -C^{\scriptscriptstyle J} du).
\end{equation*}
Here $u$ is the retarded time coordinate, $r$ is the radial luminosity coordinate 
and $x^{\scriptscriptstyle I}$ with $({\textstyle I}, {\textstyle J},\ldots)=1,2$ are the coordinates on the celestial sphere. $F, \beta$ and $C^{\scriptscriptstyle I}$ are 
functions of $u,r$ and $x^{\scriptscriptstyle I}$. $h_{{\scriptscriptstyle I} {\scriptscriptstyle J}}= \gamma_{{\scriptscriptstyle I} {\scriptscriptstyle J}} + C_{{\scriptscriptstyle I} {\scriptscriptstyle J}}/r + o(r^{-1})$ where $\gamma_{{\scriptscriptstyle I} {\scriptscriptstyle J}}$ is the metric on the 
round sphere, and $C_{{\scriptscriptstyle I} {\scriptscriptstyle J}}$ describes gravitational radiation escaping to null infinity from the interior 
of the spacetime. $C_{{\scriptscriptstyle I} {\scriptscriptstyle J}}$ is a function of $u$ and $x^{\scriptscriptstyle I} $ and $\gamma^{{\scriptscriptstyle I} {\scriptscriptstyle J}}C_{{\scriptscriptstyle I} {\scriptscriptstyle J}}=0$. It thus has two degrees
of freedom corresponding to the two possible polarization states of gravitational waves. The Bondi news tensor 
is $N_{{\scriptscriptstyle I} {\scriptscriptstyle J}}=\partial_uC_{{\scriptscriptstyle I} {\scriptscriptstyle J}}$ and is a measure of gravitational radiation, the energy flux being
$\frac{1}{32\pi}N_{{\scriptscriptstyle I} {\scriptscriptstyle J}}N^{{\scriptscriptstyle I} {\scriptscriptstyle J}}$.  Finally, $F=1-\frac{2M}{r}+o(r^{-1})$ where
 $M$ is the Bondi mass aspect. The integral of $M$ over the 2-sphere at null infinity is the Bondi mass.
 
In choosing Bondi coordinates, four degrees of freedom of the metric have been 
eliminated by setting $g_{rr}=g_{ri}=0$ and $\det(h_{{\scriptscriptstyle I} {\scriptscriptstyle J}}) = \det (\gamma_{{\scriptscriptstyle I} {\scriptscriptstyle J}})$. The residual diffeomorphisms 
that generate asymptotic symmetries are supertranslations and superrotations and their descendants,
\begin{eqnarray*}
 & \xi^u=f, \quad 
\xi^r= \frac{r}{2} \left( C^{{\scriptscriptstyle I}} \partial_{{\scriptscriptstyle I}} f - D_{{\scriptscriptstyle I}} \xi^{{\scriptscriptstyle I}} \right),\\ 
&\xi^{{\scriptscriptstyle I}} = Y^{{\scriptscriptstyle I}} - \int_{r}^{\infty} dr' \frac{e^{2\beta}}{{r'}^{2}}  h^{{\scriptscriptstyle I}{\scriptscriptstyle J}} \partial_{{\scriptscriptstyle J}} f
\end{eqnarray*}
where $f=s + \frac{u}{2} D_{\scriptscriptstyle I}Y^{\scriptscriptstyle I}$ with $s$ any spherical harmonic and $D$ is the covariant
 derivative on the unit $2$-sphere with metric $\gamma_{{\scriptscriptstyle I}{\scriptscriptstyle J}}$.
While these BMS generators are well-known, in our first order approach,  they are accompanied by Lorentz transformations that preserve the Lorentz gauge which requires six choices. We choose our basis $1$-forms
 to be $e^0=\frac{1}{2}Fdu+dr, e^1=e^{2\beta} du $ and $e^i=r E^i_{\scriptscriptstyle I}(dx^{\scriptscriptstyle I}-C^{\scriptscriptstyle I}du)$. $E^i$ is the zweibein for the metric $h_{{\scriptscriptstyle I} {\scriptscriptstyle J}}$. 
The asymptotic Lorentz translations are parametrised by
\begin{eqnarray*}
 &\Lambda_{01}= - \partial_r \xi^r, \quad \Lambda_{1 i} = \frac{1}{2r} E_{i}^{{\scriptscriptstyle I}} \left( F \partial_{{\scriptscriptstyle I}}f + 2 \partial_{{\scriptscriptstyle I}} \xi^r  \right), \\
& \Lambda_{0i} = \frac{e^{2 \beta}}{r} E_{i}^{{\scriptscriptstyle I}} \partial_{{\scriptscriptstyle I}} f, \quad
 \Lambda_{ij} = \gamma_{{\scriptscriptstyle I}{\scriptscriptstyle J}} \hat{E}^{{\scriptscriptstyle I}}_{[i} \mathcal{L}_{Y} \hat{E}^{{\scriptscriptstyle J}}_{j]} + o(r^{0})
\end{eqnarray*}
where $\hat{E}^i$ is the zweibein for the metric on the unit $2$-sphere with metric $\gamma_{{\scriptscriptstyle I}{\scriptscriptstyle J}}$.

\section{Asymptotic charges}
We are now in a position to evaluate the asymptotic charges \eqref{ham} for the Palatini and the Nieh-Yan actions. 
Assuming the fermion mass is not zero so that the fermion energy-momentum is 
 exponentially suppressed at null infinity, hence ignoring fermions, the electric charges from the Palatini action are
 \begin{equation}
 \slashed\delta \mathcal{Q} = \frac{1}{16 \pi G} \varepsilon_{abcd} \int_{\partial \Sigma} \, \iota_{\xi}e^c \, \delta \omega^{ab} \wedge e^d.\label{BMS}
\end{equation}
At leading order, they correspond to BMS charges\footnote{For brevity, we have suppressed the rotation terms involving the $Y$ generators.} (\textit{Cf.} Results in \cite{BarTro})
 \begin{equation}
\slashed\delta \mathcal{Q} = \frac{1}{16\pi}\int_{\partial\Sigma} d\Omega (4f\delta M + \frac{1}{2}fN_{{\scriptscriptstyle I}{\scriptscriptstyle J}}\delta C^{{\scriptscriptstyle I}{\scriptscriptstyle J}})
 \end{equation}
 where $d\Omega$ is the volume element on the unit $S^2$.
 The first term is integrable and is just the variation of the moments of the Bondi mass aspect $M$.
 The second term is not integrable as it is not of the form $f$ times the variation of something. 
We can identify the non-integrable term with gravitational radiation leaving the system
 and causing the mass thereby to change \cite{WZ}. Such a contribution should not be counted as part of the charge 
 on the surface as it does not describe the state of the system but rather the change of state of the system.
 We conclude that the correct expression for the change in physical charge  is just  the integrable piece
 \begin{equation}
 \mathcal{Q} = \frac{1}{4\pi} \int_{\partial\Sigma} d\Omega \ fM.
 \end{equation}
 If we ask how does this change as one goes along null infinity, we see there are two contributions \cite{bondi, sachs, Strominger:2017zoo},  
 \begin{equation}
  \partial_u M = -\frac{1}{8}N_{{\scriptscriptstyle I}{\scriptscriptstyle J}}N^{{\scriptscriptstyle I}{\scriptscriptstyle J}} + \frac{1}{4} D_{{\scriptscriptstyle I}} D_{{\scriptscriptstyle J}} N^{{\scriptscriptstyle I}{\scriptscriptstyle J}}.
  \end{equation}
The first term on the rhs is the gravitational flux, the hard component of the charge and the second is a soft component. The latter term contains contributions from soft gravitons and has physical content, for example it can be used to derive the Weinberg soft graviton theorem \cite{He:2014laa}. 

If we prescribe boundary conditions for lower orders in a $1/r$ expansion of the metric components, then we will also have subleading charges. In such a case the subleading charges obtained from \eqref{BMS} correspond to the subleading BMS charges found in \cite{fakenews}.
 
 Repeating this calculation for the Nieh-Yan action, we find that the 2-form charge from this action is equivalent to 
 \begin{equation}
\slashed\delta \mathcal{\tilde{Q}} = -\frac{1}{8\pi}\int_{\partial\Sigma}  \delta e^a \swedge \mathcal{L}_\xi e_a.
 \end{equation}
Full and further details of these charges will be 
presented in a forthcoming publication. 

The asymptotic charges that are obtained from the Nieh-Yan action correspond to dual 
charges \cite{dual0, dualex}, which at leading order are given by an integral of the NUT aspect and encodes higher moments of the topological properties of the spacetime, for example the NUT charge. 
This gives a Hamiltonian derivation of dual charges: they are the asymptotic charges that arise by 
considering the Nieh-Yan action. This is analogous to getting magnetic charges from the $\theta$-term 
in electromagnetism. However, in gravity, we see that this is only possible in a first order formalism 
and cannot be achieved in the metric formulation.  

\section{Other possible terms}

As topological terms, we should also consider the Pontryagin action $\tfrac{1}{2} \int \mathcal{R}_{ab} \swedge \mathcal{R}^{ab}$ and 
the Gauss-Bonnet action $\tfrac{1}{2}\varepsilon_{abcd}\int \mathcal{R}^{ab} \swedge \mathcal{R}^{cd} $  which, while 
higher-derivative, do not modify the Einstein equation. The equations of motion from these actions are the
differential Bianchi identity and its Hodge dual.
The presymplectic forms are
\begin{equation}
 \theta_{P} =  \delta \omega^{ab} \swedge \mathcal{R}_{ab}, \quad   \theta_{GB} = \varepsilon_{abcd} \, \delta \omega^{ab} \swedge \mathcal{R}^{cd},
\end{equation}
for the Pontryagin and Gauss-Bonnet terms, respectively. Furthermore, the Noether charges are
\begin{eqnarray}
& Q_{P} = ( \iota_{\xi}\omega^{ab} - \Lambda^{ab}) \mathcal{R}_{ab}, \\
 & Q_{GB} = \varepsilon_{abcd} \, ( \iota_{\xi}\omega^{ab} - \Lambda^{ab}) \mathcal{R}^{cd}.
\end{eqnarray}
However, as we already discussed the physical object is the asymptotic charge, coming from a Hamiltonian flow, given by equation \eqref{ham}. We can show that for the Pontryagin and Gauss-Bonnet actions respectively
\begin{eqnarray}
& \slashed\delta \mathcal{Q}_{P} = \delta \omega^{ab} \swedge \K_{\xi,\Lambda} \omega_{ab}, \\
& \slashed\delta \mathcal{Q}_{GB} = \varepsilon_{abcd} \, \delta \omega^{ab} \swedge \K_{\xi,\Lambda}\omega^{cd}, 
\end{eqnarray}
where $\K_{\xi,\Lambda}\omega^{ab}$ is
\begin{equation}
 \K_{\xi,\Lambda} \omega^{ab} = \mathcal{L}_{\xi} \omega^{ab} - d \Lambda^{ab} + [\Lambda,  \omega]^{ab}.
\end{equation}
From the asymptotic boundary conditions, we find that there is no leading order nor $O(1/r)$ asymptotic charge corresponding to the Pontryagin and Gauss-Bonnet actions. However, there are non-trivial charges at subleading $O(1/r^2).$ What these charges at subleading orders are depends on how much analyticity we allow at lower orders in the boundary conditions. Full and further details of these charges will be presented in a forthcoming publication. 

In this letter, we have argued that a full understanding of asymptotic charges in gravity requires the inclusion of all possible actions that give rise to the Einstein equation. We have shown that, for example, the Nieh-Yan term gives rise to dual charges that encode topological information about the spacetime.   

\begin{acknowledgments}
\noindent
{\textbf{Acknowledgements}} We would like to thanks Gary Gibbons and Chris Pope for discussions.  We would like to thank the Mitchell Family Foundation for hospitality at the 2019 Cook's Branch workshop and for continuing support. MG and MJP would like to thank the Max-Planck-Institut f\"ur Gravitationsphysik (Albert-Einstein-Institut), Potsdam and HG would like to thank Queen Mary University of London for hospitality during the course of this work. HG is supported by the ERC Advanced Grant “Exceptional Quantum Gravity” (Grant No. 740209).  MG is supported by a Royal Society University Research Fellowship. MJP is supported by an STFC consolidated grant ST/L000415/1, String Theory, Gauge Theory and Duality. 

\end{acknowledgments}

\bibliography{charge.bib}

\end{document}